\documentclass[letter]{aa} 
\usepackage{natbib}
\bibpunct{(}{)}{;}{a}{}{,}
\usepackage{graphicx}
\usepackage{txfonts}
\usepackage[colorlinks,citecolor=blue,urlcolor=blue,bookmarks=false,hypertexnames=true]{hyperref}
\usepackage{soul}
\begin{document} 
	
	\title{Enhanced proton parallel temperature inside patches of switchbacks in the inner heliosphere}
	
	\author{L. D. Woodham
		\inst{1} \and T. S. Horbury\inst{1} \and L. Matteini$^{1}$ \and T. Woolley$^{1}$ \and R. Laker$^{1}$ \and S. D. Bale$^{1-4}$ \and G. Nicolaou$^{5}$ \and J. E. Stawarz$^{1}$ \and D. Stansby$^{5}$ \and H. Hietala$^{1}$ \and D. E. Larson$^{2}$ \and R. Livi$^{2}$ \and J. L. Verniero$^{2}$ \and M. McManus$^{2}$ \and J. C. Kasper$^{6}$ \and K. E. Korreck$^{7}$ \and N. Raouafi$^{8}$ \and M. Moncuquet$^{9}$ \and M. P. Pulupa$^{2}$
	}
	
	\institute{Department of Physics, The Blackett Laboratory, Imperial College London, London, UK
		\and
		Space Sciences Laboratory, University of California, Berkeley, CA, USA
		\and
		Physics Department, University of California, Berkeley, CA, USA
		\and
		School of Physics and Astronomy, Queen Mary University of London, London, UK
		\and
		Mullard Space Science Laboratory, University College London, Dorking, UK
		\and
		Department of Climate and Space Sciences and Engineering, University of Michigan, Ann Arbor, MI, USA
		\and
		Smithsonian Astrophysical Observatory, Harvard-Smithsonian Center for Astrophysics, Cambridge, MA, USA
		\and
		Applied Physics Laboratory, Johns Hopkins University, Laurel, MD, USA
		\and
		LESIA, Observatoire de Paris, Universit\'e PSL, Meudon, France
	}
	
	\date{Received TBD; accepted TBD}
	
	\abstract
	{Switchbacks are discrete angular deflections in the solar wind magnetic field that have been observed throughout the heliosphere. Recent observations by \textit{Parker Solar Probe} (PSP) have revealed the presence of patches of switchbacks on the scale of hours to days, separated by `quieter' radial fields.}
	{We aim to further diagnose the origin of these patches using measurements of proton temperature anisotropy that can illuminate possible links to formation processes in the solar corona.}
	{We fitted 3D bi-Maxwellian functions to the core of proton velocity distributions measured by the SPAN-Ai instrument onboard PSP to obtain the proton parallel, $T_{p,\|}$, and perpendicular, $T_{p,\perp}$, temperature.}
	{We show that the presence of patches is highlighted by a transverse deflection in the flow and magnetic field away from the radial direction. These deflections are correlated with enhancements in $T_{p,\|}$, while $T_{p,\perp}$ remains relatively constant. Patches sometimes exhibit small proton and electron density enhancements.}
	{We interpret that patches are not simply a group of switchbacks, but rather switchbacks are embedded within a larger-scale structure identified by enhanced $T_{p,\|}$ that is distinct from the surrounding solar wind. We suggest that these observations are consistent with formation by reconnection-associated mechanisms in the corona.}
	
	
	\titlerunning{Proton Temperature in Patches of Switchbacks}
	
	\maketitle
	%
	
   \section{Introduction} \label{sec:Intro}
   
   Switchbacks in the solar wind have been observed throughout the heliosphere with increasing prominence closer to the Sun \citep[e.g.][]{Balogh1999,Neugebauer2013,Horbury2018,Owens2018,Macneil2020}. These Alfv\'enic structures often maintain a nearly constant field strength, |$\textbf{B}|$, consisting of a rapid reversal in the direction of the background radial magnetic field, $\textbf{B}_R$, as well as enhanced radial velocity, $\textbf{v}_R$, over the background flow \citep{Matteini2014,Matteini2015}. \textit{Parker Solar Probe} \citep[PSP;][]{Fox2016} observations during solar encounters have revealed the presence of patches of switchbacks on the scale of hours to days, separated by intervals of `quieter' radial fields \citep{Bale2019,Kasper2019,Horbury2020}. The origin of these structures is still poorly understood and it is not yet clear whether they result from sudden or impulsive events in the chromosphere and corona \citep{Roberts2018c,Tenerani2020,Sterling2020,Fisk2020} or are steepened waves driven by turbulence and plasma expansion \citep{Squire2020a}.
   
   Several recent studies have investigated the evolution and nature of switchbacks using PSP observations \citep{DudokdeWit2020,Farrell2020,Horbury2020,Krasnoselskikh2020,McManus2020,Mozer2020,Tenerani2020}. Probing the microphysics of these structures is essential to diagnose their origin and contribution to the total energy and momentum flux of the solar wind. Most PSP observations to date constitute a reduced 1D velocity distribution function (VDF) along the radial direction, making the investigation of the thermodynamic properties of switchbacks difficult \citep[e.g.][]{Huang2020,Mozer2020}. Recently, \citet{Woolley2020} identified individual switchbacks with a full 180$^\circ$ rotation in $\textbf{B}$ to show that the parallel temperature inside the structures is similar to the outside value. \citet{Verniero2020} present the first analysis of 3D proton VDFs inside and outside a switchback at 35 solar radii, using 1D fitting to characterise the core and proton beams. They find that the temperature of both components remained unchanged through the field reversal.
   
   In this letter, we investigate how the proton temperature varies on larger scales across patches of multiple switchbacks. We fitted a 3D bi-Maxwellian model to the core of proton VDFs measured by PSP inside 40 solar radii. We find that the presence of patches of switchbacks is correlated with enhancements in the proton parallel temperature, $T_{p,\|}$, while the perpendicular temperature, $T_{p,\perp}$, remains relatively constant. These patches are highlighted by a transverse deflection of the flow and magnetic field away from the radial direction, as well as a small increase in both proton and electron density. Individual switchbacks are embedded within these larger-scale regions of enhanced $T_{p,\|}$, indicating a possible common origin in the corona. This result is the first direct evidence of a robust increase in proton temperature inside patches, providing a possible direct link between switchbacks and their origin in the solar atmosphere.
   
   \section{Data analysis} \label{sec:Method}
   
   The SWEAP suite of instruments on PSP consists of several electrostatic analysers (ESA) and a Faraday cup \citep{Kasper2016}. SPAN-Ai is a top-hat ESA \citep{Carlson1982} located on the ram side of the spacecraft that exclusively measures ions, including a time-of-flight recorder to differentiate between species such as protons and $\alpha$-particles \citep{Livi2020}. During a solar encounter, the heat shield of the spacecraft partially obscures the field-of-view of SPAN-Ai and the measured ion VDFs are cut off in the plane tangential to the the spacecraft trajectory. To compensate for this, the Solar Probe Cup \citep[SPC;][]{Case2020} looks directly around the heat shield and radially towards the Sun, measuring a reduced 1D distribution function. The combination of SPC and SPAN-Ai measurements allows for a near-full determination of ion distributions at perihelion close to the Sun. In this letter, we primarily use SPAN-Ai data, fitting a model bi-Maxwellian to 3D proton VDFs at $\sim$7 s cadence. We complement these fits with data from SPC at $\sim$0.218 s resolution. We also use magnetic field measurements from the FIELDS fluxgate magnetometer \cite[MAG;][]{Bale2016}, downsampled to 16 vectors/s, as well as electron density, $n_e$, derived from quasi-thermal noise (QTN) measurements by the Radio Frequency Spectrometer \citep[RFS;][]{Pulupa2017,Moncuquet2020}.
   
   \begin{figure}
   	\centering
   	\includegraphics[width=0.9\linewidth]{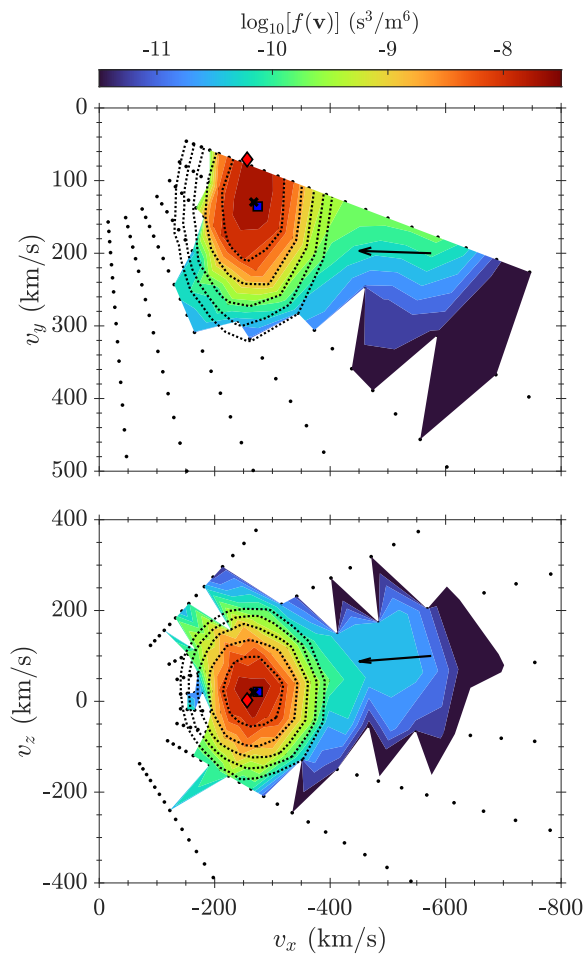} 
   	\caption{Example proton VDF (5 April 2019 20:21:36.7407) in the SPAN-Ai instrument frame. The dotted contours are the bi-Maxwellian fit to the proton core using Equation \ref{eq:modelf} with $\mathbf{u}$ indicated by the black cross. The blue square is the bulk velocity moment of the SPAN-Ai distribution and the red diamond is the average proton core bulk velocity measured by SPC during the SPAN-Ai integration time. The black arrow gives the average direction of $\textbf{B}$ during this time.}
   	\label{fig:1}
   \end{figure}
   
   Proton VDFs measured by SPAN-Ai often show the presence of both a proton core and a field-aligned beam; a second proton component that streams faster than the core along the direction of $\hat{\textbf{b}}=\textbf{B}/|\textbf{B}|$ \citep[see][]{Verniero2020}. In Figure \ref{fig:1}, we show a typical distribution measured by SPAN-Ai in the instrument frame\footnote{See Appendix \ref{App:A} for details on coordinate systems used in this letter.}. The limited field-of-view (FOV) of the instrument is apparent in the top panel, where both the core and beam are cut off by the spacecraft heat shield; the latter of which is almost completely obscured due to the orientation of the magnetic field. As the solar wind flow in the instrument frame deflects further in the $\hat{\textbf{y}}$-direction, the proton distribution shifts further into the instrument FOV. This is achieved either by a physical deflection in the flow or by increasing spacecraft velocity.
   
   \begin{figure*}
   	\centering
   	\includegraphics[width=0.95\linewidth]{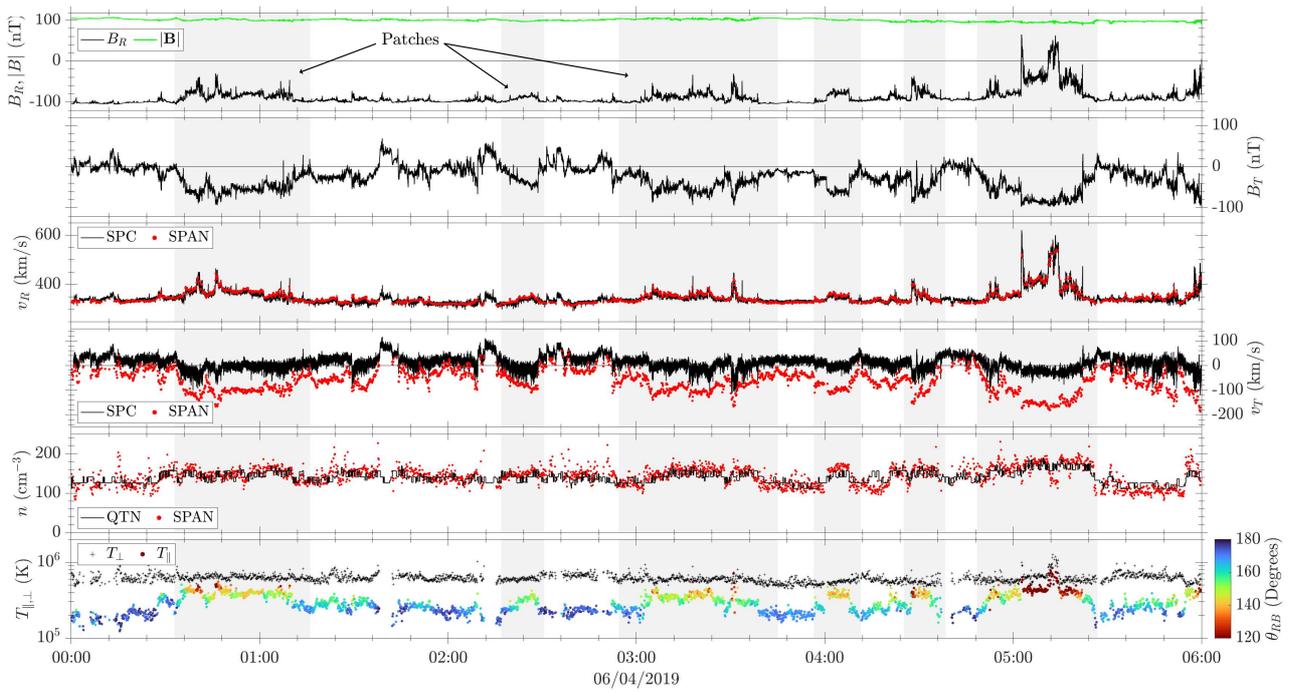}
   	\caption{Interval of PSP observations during Encounter 2. First panel: radial component of the magnetic field, $B_R$, and the field magnitude, $|\textbf{B}|$. Second panel: tangential component of the field, $B_T$. Third and fourth panels: radial and tangential components of the proton core velocity, $v_R$ and $v_T$, respectively. Here, the black lines are the measurements of the proton core by SPC \citep[for details, see][]{Case2020} and the red dots are the fits to the proton core from SPAN-Ai. All measurements are in RTN coordinates. Fifth panel: proton core density, $n_p$, from SPAN and electron density, $n_e$, from QTN measurements. Last panel: proton parallel ($T_{p,\|}$) and perpendicular ($T_{p,\perp}$) temperature, where the colour-scale of $T_{p,\|}$ is the angle, $\theta_{RB}$, between the radial direction and $\textbf{B}$. During this interval the large-scale field is sunward so that $\theta_{RB}=180^\circ$ indicates the radial direction.}
   	\label{fig:2}
   \end{figure*}
   
   In this study, we focus only on the proton core. To obtain proton bulk parameters, we first transformed the distribution from the instrument frame $(v_x,v_y,v_z)$ into field-aligned coordinates $(v_\|,v_{\perp1},v_{\perp2})$ using the Euler-Rodrigues formula \citep[for more details, see][]{Valdenebro2016}. The rotation matrix for this coordinate transformation is:
   
   \begin{equation}
   	\mathsf{T}=\left(\begin{array}{ccc}
   		\cos \phi & -k_{z} \sin \phi & k_{y} \sin \phi \\
   		k_{z} \sin \phi & k_{y}^{2}+k_{z}^{2} \cos \phi & k_{y} k_{z}(1-\cos \phi) \\
   		-k_{y} \sin \phi & k_{y} k_{z}(1-\cos \phi) & k_{z}^{2}+k_{y}^{2} \cos \phi
   	\end{array}\right),
   \end{equation}
   
   \noindent where $\hat{\textbf{k}}=(\hat{\textbf{x}}\times\hat{\textbf{b}})/|\hat{\textbf{x}}\times\hat{\textbf{b}}|$ is the unit-vector along the rotation axis, $\hat{\textbf{x}}=(1,0,0)$ is the axis of the instrument frame directed towards the Sun, and $\phi$ is the angle between $\hat{\textbf{x}}$ and $\hat{\textbf{b}}$. Here, the coordinate frame is rotated by $\phi$ about the axis defined by $\hat{\textbf{k}}$. Therefore, the $v_{\perp,1}$ and $v_{\perp,2}$ directions are defined with respect to the instrument coordinate system and the rotation axis, as opposed to a heliographic direction. We performed a non-linear least squares fit by minimising the sum \cite[see also][]{Bercic2019,Bercic2020,Durovcova2019,Stansby2018b,Stansby2018f,Nicolaou2020,Nicolaou2020a}:
   
   \begin{equation}
   	\chi=\sum_{i=1}^N{\left[\log_{10}{(f_{\textrm{model},i})}-\log_{10}{(f_{\textrm{meas},i})}\right]^2},
   \end{equation}
   
   \noindent for $N$ fitting points. We used a 3D model bi-Maxwellian, assuming gyrotropy in $\textbf{v}_\perp$:
   
   \begin{equation} \label{eq:modelf}
   	f(\textbf{v})=\frac{n}{\pi^{3/2}w_\perp^2 w_\|}\exp{\left(-\frac{(v_\|-u_\|)^2}{w_\|^2}-\frac{(\mathbf{v_{\perp}}-\mathbf{u_{\perp}})^2}{w_\perp^2}\right)},
   \end{equation}
   
   \noindent where $n$ is the number density, $\textbf{u}=(u_\|,u_{\perp1},u_{\perp2})$ is the bulk velocity, and $w_\|$ and $w_\perp$ are the thermal speeds parallel and perpendicular to $\hat{\textbf{b}}$, respectively. The thermal speed is related to the temperature by $w_{\|,\perp}=\sqrt{2k_BT_{\|,\perp}/m}$, where $m$ is the proton mass. We include our bi-Maxwellian fit to the proton core in Figure \ref{fig:1}, showing good agreement with the measured distribution.
   
   We obtain proton core parallel and perpendicular temperature measurements for times when the distributions are not obscured by the spacecraft heat-shield. To retain a fit, we required at least three bins in the $\hat{x}$-$\hat{y}$ plane. This ensures that enough of the proton core is visible to the SPAN-Ai FOV in order to identify the centre of the peak accurately. We also quantified the angular fluctuations in $\textbf{B}$ during the distribution integration time with:
   
   \begin{equation} \label{eq:devB}
   	\psi_{B}=\frac{1}{N}\sum_{i=1}^{N} \arccos \left(\hat{\mathbf{b}}_{i} \cdot \hat{\mathbf{b}}_{\textrm{SPAN}}\right),
   \end{equation}
   
   \noindent where $\hat{\mathbf{b}}_{\textrm{SPAN}}$ is the average magnetic field direction over the $\sim$7 s measurement interval, $\hat{\mathbf{b}}_{i}$ is the instantaneous magnetic field unit-vector, and $N$ is the number of measurements. Large fluctuations in $\textbf{B}$ over the integration time result in a broadening of the VDFs that increases uncertainty in the measurements of proton temperature anisotropy \citep[e.g.][]{Verscharen2011}. To reduce this blurring effect, we excluded fits with angular deviations $\psi_B>10^\circ$. Finally, we manually removed times when the proton beam became so large that an automated determination of the core was not possible. We also did this by eye for any fits that we deemed physically unreasonable.
   
   \section{Results} \label{sec:Results}
   
   In Figure \ref{fig:2}, we present an example six-hour interval from after perihelion during PSP's second encounter. During this encounter, PSP primarily measured slow Alfv\'enic wind with complicated solar source mapping \citep{Rouillard2020}. The properties of slow Alfv\'enic wind and possible coronal source regions have been studied extensively \citep{DAmicis2015,DAmicis2019,Huang2020c,Stansby2019,Stansby2020a,Perrone2020}. We see the presence of several patches indicated by the shaded regions. These structures exhibit characteristic negative deflections in both $B_T$ and $V_T$. This diagnostic of $B_T,v_T\ne0$ implies that patches are not just groups of switchbacks, which are easily identified in $B_R$, but rather part of a larger-scale structure that are cut through by the spacecraft. The field magnitude, $|\textbf{B}|$, is also relatively constant across this interval, indicating that these structures are rotations of the magnetic field vector about a sphere of constant radius \citep{Matteini2014,Matteini2015}.  Multiple magnetic switchbacks are embedded inside each patch where the radial field, $B_R$, increases rapidly, indicating a rotation in the field. These switchbacks exhibit different angular rotations over a range of timescales and they appear superposed on the background radial field. We also see similar enhancements in $v_R$ inside switchbacks due to their Alfv\'enic nature. In this interval, the majority of switchback structures do not rotate more than 90$^\circ$ and last for tens of minutes.
   
   There is good agreement between SPC and SPAN-Ai in Figure \ref{fig:2} for the radial component for the velocity, but this is not the case for $v_T$. Instead, we see that while SPC measurements largely follow the variation seen in SPAN-Ai data, they underestimate the magnitude of the flow deflection from the radial direction. This is because for large $-v_T$ flows, the solar wind flow is at a large angle of incidence into the cup, so that a radial cut through the distribution likely captures only the wings of the proton core \citep{Kasper2016,Case2020}. As the flow returns to radial into the cup (i.e. $v_T$ $\sim$ 80 km/s, accounting for the spacecraft velocity), SPC measurements become more reliable and both instruments are in good agreement. As $v_T>0$, the proton core shifts too far out of the SPAN-Ai FOV and measurements become unreliable. Our fitting of the proton VDFs here reveals larger negative tangential flows than previously reported with SPC \citep{Kasper2019}, with implications for flow circulation in the solar atmosphere \citep{Fisk2020}.
   
   \begin{figure}
   	\centering
   	\includegraphics[width=0.95\linewidth]{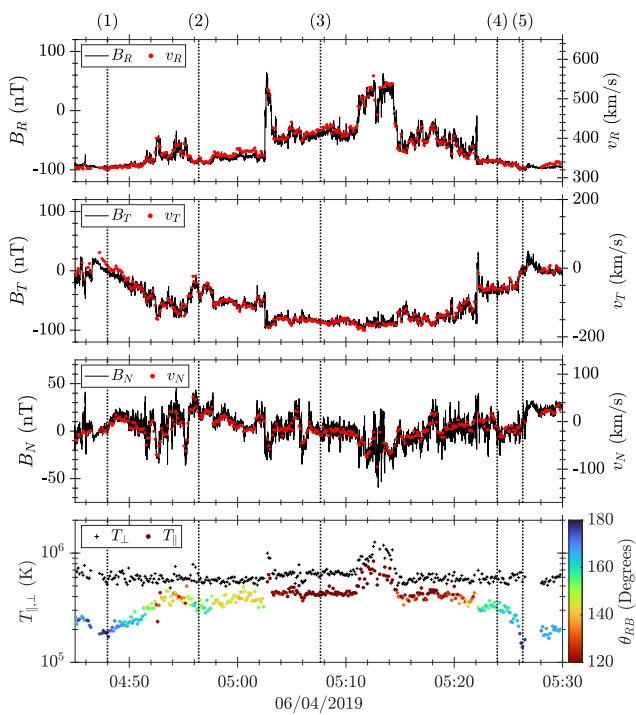}
   	\caption{Single patch from PSP observations. First panel: radial component of the magnetic field, $B_R$, and proton core velocity, $v_R$. Second and third panels: same as the first panel, but for the tangential and normal components, respectively. Last panel: $T_{p,\|}$ and $T_{p,\perp}$, where the colour-scale of $T_{p,\|}$ is the angle $\theta_{RB}$. The vertical dashed lines indicate the times of the distributions plotted in Figure \ref{fig:4}.
   	}
   	\label{fig:3}
   \end{figure}
   
   \begin{figure*}
   	\centering
   	\includegraphics[width=0.8\linewidth]{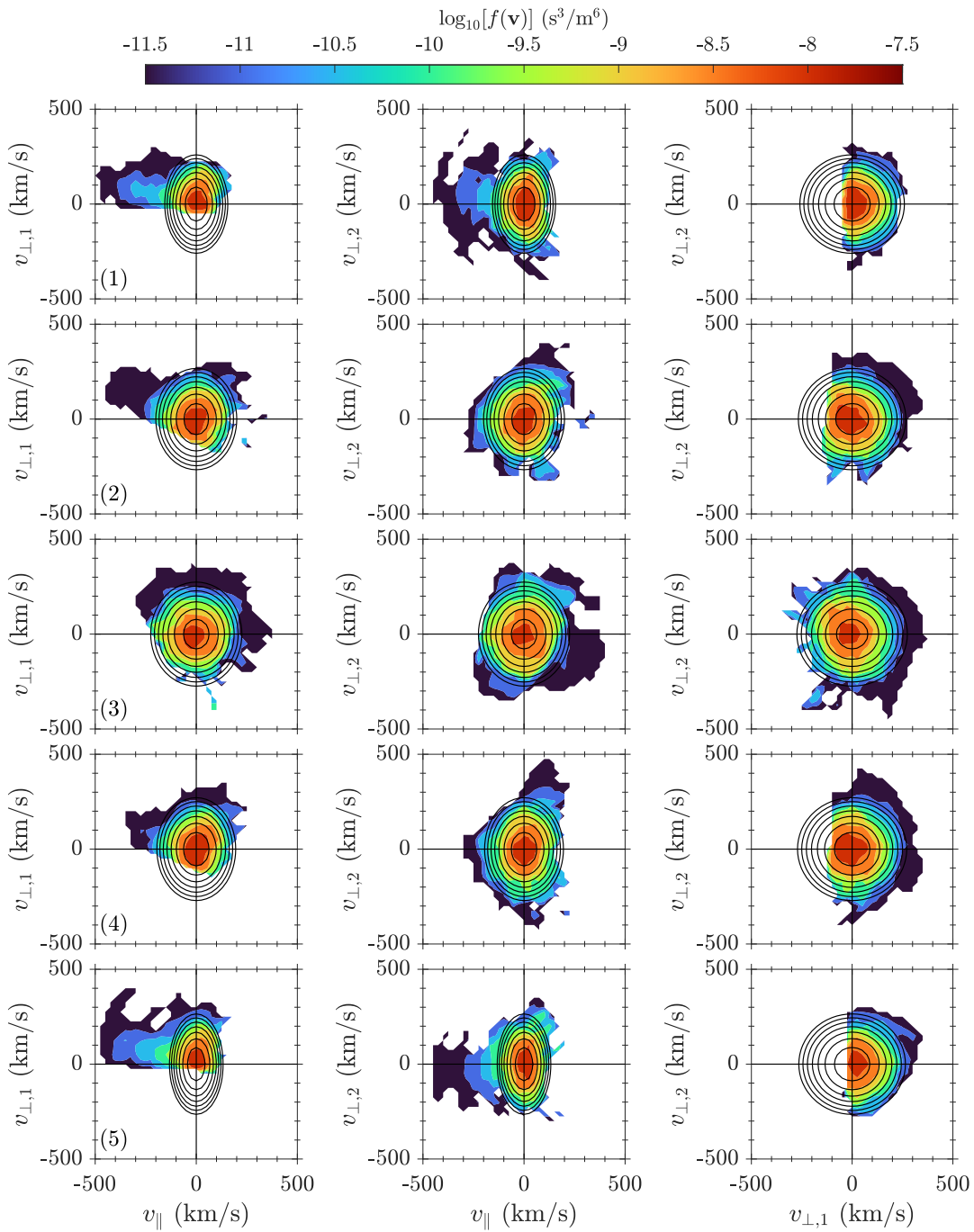} 
   	\caption{Examples of proton distributions measured by SPAN-Ai during the patch shown in Figure \ref{fig:3} where each row corresponds to the times indicated. Each column gives a cut of the measured proton distribution in the proton core bulk frame, i.e. the core is centred on $(0,0)$. For example the left column is a cut through the distribution in the $v_\|$-$v_{\perp,1}$ plane at $v_{\perp,2}=0$. We also include our bi-Maxwellian fits as contours, showing overall good agreement with the distributions.}
   	\label{fig:4}
   \end{figure*}
   
   From the bottom panel in Figure \ref{fig:2}, we see that $T_{p,\perp}$ is approximately constant across the entire interval. In contrast, $T_{p,\|}$ exhibits a systematic variation that correlates with the presence of the patches seen as deflections in both $B_T$ and $v_T$. Under spherical polarisation with constant $|\textbf{B}|$, $B_T$ and $v_T$ are mathematically related to $\theta_{RB}$ and they correlate with each other. We see that $T_{p,\|}$ is highly anti-correlated with both $B_T$ and $v_T$, resulting in a dependence on $\theta_{RB}$. These enhancements in $T_{p,\|}$ vary both with the presence of individual switchbacks as well as the large-scale structure of a single patch. This combined plasma and magnetic field signature of a patch is the main result of this letter. Overall, $T_{p,\perp}>T_{p,\|}$ throughout the interval, which is consistent with many observations of fast wind in the inner heliosphere \citep{Durovcova2019,Hellinger2011,Hellinger2013,Marsch1981,Marsch1982b,Marsch2004,Matteini2007,Perrone2018,Perrone2019,Stansby2019,Stansby2018f,Stansby2020a} and the recent analysis of slow Alfv\'enic wind observed by PSP \citep{Huang2020c,Huang2020,Verniero2020}. In addition, we see in panel (5) that patches sometimes display small density enhancements in both $n_p$ and $n_e$. Here we normalise $n_p$ from SPAN-Ai in the figure to the mean value of $n_e$ from QTN. Refinements to the calibration of the instrument are on-going and so we do not focus on the magnitude of $n_p$ here. Despite this, we see that the variation of both $n_p$ and $n_e$ inside patches are in good agreement.
   
   To investigate the enhancement in $T_{p,\|}$ further, we plotted a shorter interval of a single patch in Figure \ref{fig:3}. This patch is characterised by a clear deflection away from $B_T=v_T=0$, while $B_R$ and $v_R$ show the presence of multiple smaller-scale switchbacks. We see no distinct large-scale structure of the patch in either $B_N$ or $v_N$, despite deflection inside individual switchbacks. At around 05:15, there is a switchback with a defection in the normal direction while $v_T$ remains roughly constant. This switchback is embedded within a larger switchback structure, and it exhibits an increase in both $T_{p,\|}$ and $T_{p,\perp}$ that is not seen elsewhere in Figure \ref{fig:2}. Throughout the rest of the patch, there is a clear increase in $T_{p,\|}$ while $T_{p,\perp}$ remains constant. We see the anti-correlation of $T_{p,\|}$ again with $B_T$ and $v_T$ in panels (2) and (4). There is also a clear dependence of $T_{p,\|}$ on $\theta_{RB}$, which is seen over the large scale patch structure as well as inside individual switchbacks, for example, at 05:03. In general, as the magnetic field begins to rotate away from the radial direction, there is a large increase in $T_{p,\|}$ that begins to saturate at a maximum value of $\sim$4$\times10^5$ K as the field continues to deflect up to 60$^\circ$ from the radial direction. As the field vector returns to the radial direction, $T_{p,\|}$ also returns to the background value of $\sim$2$\times10^5$ K outside the patch.
   
   We analyse whether the limited FOV of SPAN-Ai leads to an artificial enhancement in $T_{p,\|}$ by plotting cuts of the measured proton distributions across the patch in Figure \ref{fig:4}. We took cuts at $v_\|=0$, $v_{\perp,1}=0$, and $v_{\perp,2}=0$ in the proton core bulk frame and included our bi-Maxwellian fit to each distribution. We see from the top and bottom rows that the distributions at the edges of the patch show clear anisotropy with $T_{p,\perp}>T_{p,\|}$. The magnetic field is radial at these times and so the presence of a proton beam is seen in the SPAN-Ai FOV due to its location in velocity space, although these features are more obscured than the core. As PSP moves deeper inside the patch in rows (2) and (3), more of the distribution is seen as the solar wind flow is defected in $-v_T$. However, as the field also rotates in the $-\hat{\mathbf{T}}$ direction, the beam moves out of the instrument FOV. We also see a clear broadening of the proton core along $\textbf{B}$ so that $T_{p,\perp}/T_{p,\|}$ approaches unity. As the spacecraft measures the trailing edge of the patch, row (4) shows a decrease in $T_{p,\|}$ back to a similar value before the patch. These distributions corroborate our main result, showing that while the proton core may be partially obscured outside of a patch, there is a clear broadening of the distribution parallel to $\textbf{B}$ inside the patch. In fact, since the field is typically radial outside of a patch, we would expect more uncertainty in $T_{p,\perp}$ compared to $T_{p,\|}$, as seen by the clear cut-off along $v_{\perp,1}$ in rows (1) and (5). Therefore, we conclude that the  observed increase in $T_{p,\|}$ is reliable and does not simply reflect a limited FOV effect.
   
   \section{Discussion \& conclusions} \label{sec:Discussion}
   
   In this letter we perform a full 3D fitting of bi-Maxwellian functions to the core of proton velocity distributions measured by SPAN-Ai. We obtain proton temperature anisotropy while accounting for FOV limitations during the spacecraft's second encounter with the Sun. We reveal that patches of Alfv\'enic switchbacks correlate with enhancements in the proton parallel temperature, $T_{p,\|}$, while the perpendicular temperature, $T_{p,\perp}$, is consistently larger and remains relatively constant. This increase in the width of the proton distribution parallel to $\textbf{B}$ is robust, confirming that our result is not due to the systematic effect of the distribution moving into the instrument FOV. We also find that generally there is a small increase in both proton and electron density inside patches, although this is not always the case. These patches can be identified by a transverse deflection of the flow and magnetic field away from the radial direction. This result constitutes the first clear identification of a plasma signature of patches of switchbacks in the solar wind, and it provides clues as to the origin of these structures.
   
   \citet{Woolley2020} recently showed that $T_{p,\|}$ remains approximately constant in switchbacks with a deflection of $180^\circ$ using radial temperature measurements from SPC. They interpret this result to be consistent with a velocity space rotation of the plasma VDF. In contrast, we find a clear increase in $T_{p,\|}$ inside both switchbacks and patches. One possible functional form for this dependence of $T_{p,\|}$ on $\theta_{RB}$ that is consistent with both studies is $\Delta T_{p,\|}\sim\sin(\theta_{RB})$. However, at present, we have insufficient data to confirm this or any other particular dependence unambiguously using SPAN-Ai observations. As the spacecraft tangential velocity increases in future encounters, the distribution will move further into the instrument FOV, providing the opportunity to further investigate this relationship. While we do not show it here, the correlation between $T_{p,\|}$ and $\theta_{RB}$ is persistent throughout the entire second encounter, suggesting that this plasma signature is a widespread phenomenon associated with the majority of switchback structures.
   
   Our results reveal that patches are not simply a group of switchbacks, but rather switchbacks are embedded within a larger-scale structure identified by distinct plasma signatures. As suggested by previous studies \citep[e.g.][]{Horbury2018,Horbury2020}, we hypothesise that patches may be the in situ manifestation of spatially extended coronal plumes \citep{Raouafi2007}, which are cut through by PSP at $\sim$35 solar radii. Individual switchbacks may then result from intermittent reconnection outflows due to the footprint motion of the coronal magnetic field \citep[see][and references therein]{Raouafi2016}, resulting in coronal jets and jetlets \citep{Sterling2020}. However, other mechanisms cannot be ruled out, such as steepened Alfv\'en waves generated within the reconnection exhaust itself \cite[e.g.][]{Squire2020a}. Enhanced proton parallel temperatures may result from ion-scale processes within reconnection exhausts \citep[e.g.][]{Drake2009,Hietala2015,Hietala2017}. Alternatively, enhanced turbulence within switchbacks \citep{DudokdeWit2020} may lead to increased dissipation associated with smaller-scale coherent structures such as current sheets \citep[e.g.][]{Chasapis2018,Karimabadi2013,Wan2012}. In fact, \citet{Woodham2020} recently interpreted the link between enhancements in $T_{p,\|}$ at 1 AU and deflections in $\theta_{RB}$ to turbulent dissipation.
   
   One important caveat to our interpretations is the effect of the spacecraft cutting through different plasma structures \citep[e.g.][this issue]{Laker2020} on plasma measurements. Further work is ongoing to disentangle temporal and structural changes in the plasma measurements and investigate the link between reconnection signatures and patches of switchbacks in the inner heliosphere. Future coordination with remote sensing observations from Earth and the recently launched \textit{Solar Orbiter} spacecraft \citep{Muller2020} will significantly aid our understanding of intermittent solar wind sources.
   
   \begin{acknowledgements}
   	LDW and TSH are supported by the STFC consolidated grant ST/S000364/1. TW is supported by STFC grant ST/N504336/1 and RL by an Imperial College President’s scholarship. SDB acknowledges the support of the Leverhulme Trust Visiting Professor programme. The FIELDS and SWEAP teams acknowledge support from NASA contract NNN06AA01C. \textit{Parker Solar Probe} was designed, built, and is now operated by the Johns Hopkins Applied Physics Laboratory as part of NASA’s Living with a Star (LWS) program. Support from the LWS management and technical team has played a critical role in the success of the mission. All data were obtained from the \href{http://spdf.gsfc.nasa.gov}{SPDF web-site}.
   \end{acknowledgements}

%
%

\bibliographystyle{aa}
\bibliography{Bib}

\begin{appendix}
	
	\section{Coordinate systems} \label{App:A}
	
	\begin{figure}
		\centering
		\includegraphics[width=0.8\linewidth]{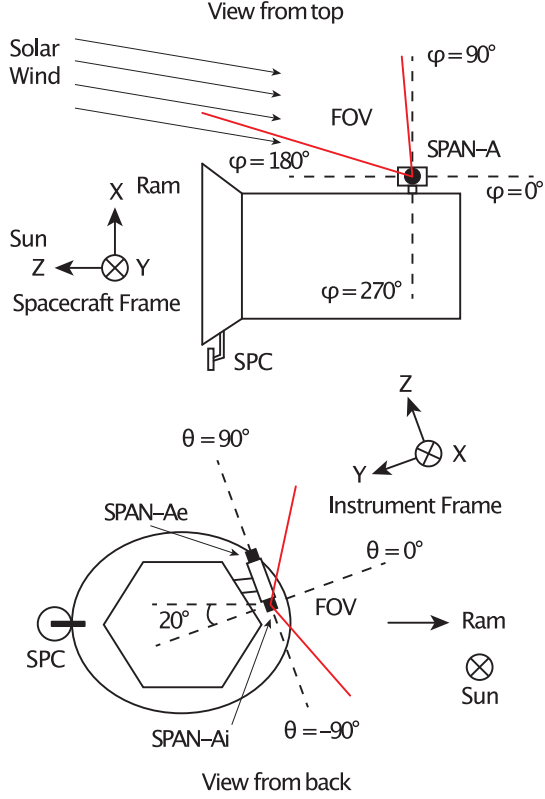} 
		\caption{Schematic of the PSP spacecraft body with orientation of the SPC and SPAN-Ai instruments and their respective coordinate frames.}
		\label{fig:A1}
	\end{figure}
	
	In Figure \ref{fig:A1}, we show a schematic of the PSP spacecraft body. SPAN-Ai is located on the ram side of the spacecraft body, whereas SPC is situated on the anti-ram side and pointed directly towards the Sun. SPAN-Ai measures particles in a spherical coordinate system: $(\phi,\theta,E)$, where $\phi$ is the azimuthal angle, $\theta$ is the elevation angle, and $E=1/2mv^2$ is the particle kinetic energy. The instrument has eight angular bins in both $\phi$ and $\theta$, with an angular coverage of 247.5$^\circ$ in azimuth and 120$^\circ$ in elevation \cite[for more details, see][]{Kasper2016,Livi2020}. The instrument is orientated such that the FOV is obstructed by the spacecraft heat-shield, blocking the line-of-sight of the Sun-spacecraft line. Therefore, the measured ion VDFs are partially obscured in the plane, tangential to the the spacecraft trajectory. To convert between a spherical and Cartesian coordinate frame, we used:

	\begin{equation}
		v_x=v\cos{\theta}\cos{\phi};\\
		v_y=v\cos{\theta}\sin{\phi};\\
		v_z=v\sin{\theta},
	\end{equation}
	
	\noindent where $\hat{\mathbf{x}}$ is directed towards the Sun, $\hat{\mathbf{y}}$ is at an angle of 20$^\circ$ to the anti-ram direction, and $\hat{\mathbf{z}}$ completes the right-handed triad. We refer to this coordinate system as the SPAN-Ai instrument frame. In contrast, SPC measurements are made in the spacecraft frame, where $\hat{\mathbf{z}}$ is directed towards the Sun, $\hat{\mathbf{x}}$ is in the ram direction, and $\hat{\mathbf{y}}$ completes the triad. The rotation matrix to convert from the SPAN instrument to the spacecraft (SPC) frame is:
	
	\begin{equation}
		\mathsf{T}=
		\begin{pmatrix}
			0 & -\cos{20^\circ} & -\sin{20^\circ} \\
			0 & \sin{20^\circ} & -\cos{20^\circ} \\
			1 & 0 & 0
		\end{pmatrix}.
	\end{equation}
	
	\noindent In this letter, we present our fitting results in the RTN coordinate system. Here, $\hat{\mathbf{R}}$ is the unit vector from the Sun towards the spacecraft, $\hat{\mathbf{T}}$ is the cross-product of the solar rotation axis and $\hat{\mathbf{R}}$, and $\hat{\mathbf{N}}$ completes the right-handed triad. During normal solar encounter orientation, the transform from spacecraft coordinates to the RTN frame are approximately as follows: $\hat{\textbf{R}}=-\hat{\textbf{z}}$, $\hat{\textbf{T}}=\hat{\textbf{x}}$, and $\hat{\textbf{N}}=-\hat{\textbf{y}}$. Since measurements taken in the SPAN or SPC frames include the velocity of the spacecraft, converting to an inertial RTN frame requires subtraction of this velocity. We performed this operation using the SPICE software package \citep{Acton2018}.
	
\end{appendix}

\end{document}